\documentclass[journal=jpcbfk,manuscript=article]{achemso}
\usepackage{graphicx}
\usepackage{epstopdf}

\author{Claudio Melis}
\affiliation[University of Cagliari]{ Dipartimento di Fisica Universit\`a di Cagliari}
\alsoaffiliation[CNR-IOM SLACS]{
Istituto Officina dei Materiali del CNR, Unit\`a SLACS, Cittadella Universitaria, I-09042 Monserrato (Ca), Italy }
\email{claudio.melis@dsf.unica.it}

\author{Alessandro Mattoni}
\affiliation[CNR-IOM SLACS]{
Istituto Officina dei Materiali del CNR, Unit\`a SLACS, Cittadella Universitaria, I-09042 Monserrato (Ca), Italy }
\email{alessandro.mattoni@dsf.unica.it}

\author{Luciano Colombo}
\affiliation[University of Cagliari]{ Dipartimento di Fisica Universit\`a di Cagliari}
\alsoaffiliation[CNR-IOM SLACS]{
Istituto Officina dei Materiali del CNR, Unit\`a SLACS, Cittadella Universitaria, I-09042 Monserrato (Ca), Italy }

\title{Self assembling of  Poly(3-hexylthiophene) (P3HT).  }

\begin{document}
\begin{abstract}
We study the assembling of P3HT chains in vacuo by means of a combination of first principles density functional theory and model potential molecular dynamics.
We find that, in the absence of any external constraints, the $\pi-\pi$ interchain interaction between thiophenes  is the major driving force for the assembling.
Single chains stack in a staggered geometry giving rise to the formation of two-dimensional hydrophobic foils. These, in turn,
assemble into a zig-zag bulk polymer structure in agreement with experimental findings. Finally, in  the presence of some external constraint (like e.g. a substrate), when the alignment of single chains is favored instead of the stacking,  a different  bulk structure is possible where  thiophene rings are aligned.
  \end{abstract}

\section{Introduction}
Semiconducting polymers are promising materials for a variety of applications in organic electronics \cite{POL-DIODES,POL-TFT} 
due to their unique combination of opto-electronic properties, ease of fabrication, and low cost  manufacturing.
Among  them, poly-3-hexylthiophene (P3HT) is one of the most extensively studied conjugated polymer 
because of its relatively high carrier mobility   (0.1 cm$^2$ V/s) coupled with solution processability\cite{P3HT-COND,P3HT-COND1,P3HT-COND2} .
P3HT applications run from light-emitting diodes\cite{P3HT-OLED} to thin-film transistors\cite{P3HT-TFT}. Moreover, P3HT coupled with inorganic materials 
(e.g. TiO$_2$\cite{MESO,CT}, ZnO\cite{gunes}, carbon nanotubes\cite{P3HT-CNT}, phenyl-C61-butyric acid methyl ester\cite{P3HT-PCBM}) is an  excellent component 
for use in low-cost photovoltaics. P3HT-based solar cells have reached power conversion efficiencies of about 5$\%$\cite{P3HT-PCBM}.

It was observed that when cast from solvents into thin films, P3HT 
self-assembles into oriented microcrystalline domains (10-50 nm) and amorphous regions\cite{P3HT-CAST,P3HT-CAST1}.
The larger the crystallinity,  the higher is the carrier delocalization, as well as mobility.
According to X-ray diffraction experiments \cite{P3HT-CAST,P3HT-CAST1,P3HT-XRAYS,P3HT-XRAYS2,P3HT-XRAYS3,P3HT-XRAYS4,P3HT-XRAYS5}
several  P3HT equilibrium crystal structures  have been proposed. 
However,  the morphological disorder occurring in polymeric systems,
mainly  due to the presence of multiple phases, makes the data interpretation a difficult task.
Crystalline P3HT  consists of a lamellar structure
where the planar polymer molecules are stacked on top of each
 other.
 Two different stacking configurations have been proposed, namely:
the "aligned"  
one\cite{P3HT-XRAYS1}  where the thiophene rings of two stacked molecules are perfectly aligned on top
of each other and  the
 "staggered" one\cite{P3HT-CAST},  where the molecules  are shifted by  
one thiophene unit   along the backbone  (\ref{P3HT-STRUCT}). 
Besides the chain stacking, also the relative orientation of molecules belonging to different stacks
is a matter of debate and several models have been proposed.
In particular, in the configuration hereafter referred to as "parallel", the hexyl  chains and the thiophene rings are  parallel\cite{P3HT-MD0,P3HT-XRAYS5}, while in the so called  "zigzag"   the hexyl  chains and the thiophene rings  are guessed to form a sort of zigzag structure \cite{P3HT-XRAYS}.

The detailed  knowledge of the polymer structure is  fundamental since the actual atomic configuration strongly affects the  
conduction properties: for instance,
it has been shown that the aligned configuration favors the interchain hole mobility\cite{P3HT-MD0}, while the intrachain mobility along the backbone is strongly improved by the chain planarity\cite{P3HT-DFT3}.

In the last few years several atomistic simulations have been performed on P3HT  in order to help the interpretation 
of the  X-ray diffraction experiments\cite{P3HT-DFT,P3HT-DFT1,P3HT-DFT2,P3HT-MD,P3HT-MD1}.
First principles density functional theory (DFT) calculations\cite{P3HT-DFT} have been applied to compare the energetics of 
the ideal staggered and aligned configurations at zero temperature. 
In particular, it was observed that  the staggered configuration is slightly more stable  
with respect to the aligned one, mainly due to a favorite spatial distribution of the side chains.
Nevertheless the calculated energy differences between the two phases at zero temperature 
are quite small  (0.45 eV/monomer) and different phases (as well as relative stability) can be expected at finite
temperature as a result of entropic contributions (e.g. the disorder of the hexyl chains).
As a matter of fact, a new polymer structure (where the hexyl chains are tilted with respect to the plane of the thiophenes)  was identified  
 by  model potential molecular dynamics (MPMD) at  room temperature\cite{P3HT-MD0}.

In this work we explore an alternative theoretical approach based on a combination of DFT and MPMD,
in which the P3HT bulk structure is  assembled in vacuo at finite temperature from
its molecular constituents. 
We study  in this way the dominant assembling mechanisms and the resulting P3HT structures.
In particular,  we identify and thoroughly characterize a new structure which is similar to experimental results \cite{P3HT-XRAYS}.

 \section{Theoretical framework}
We combine DFT and MPMD to study  both the equilibrium structure of P3HT at T=0 K and the polymer assembling at finite temperature. 

The  DFT calculations have been performed  by considering an orthorombic unit cell containing two  stacked main chains disposed either in the 
aligned or staggered configuration having in total 100 atoms each (see \ref{P3HT-STRUCT}). 
The geometries were fully relaxed with the aligned inversion iterative subspace method (DIIS)\cite{diis} by using the CPMD code\cite{cpmd} and 
norm conserving Martins-Troulliers pseudopotentials\cite{Martin}. The electronic wavefunctions were expanded in a plane wave basis set 
with a kinetic energy cutoff of 70 Ry. The exchange correlation energies were calculated using the LDA  functional\cite{LDA}. 
The results obtained with the CPMD code were further validated by performing the same calculations using the Quantum Espresso package\cite{ESPRESSO}
on a limited number of selected structures. The results were very similar with both codes.

All the configurations relaxed with DFT were further optimized by means of MPMD with a minimization procedure combining both 
local conjugate gradients minimization and low temperature molecular dynamics as described elsewhere\cite{Melis}. 
Simulations were performed with the DL\_POLY\cite{dlpoly}(version 2.19) code by using  the AMBER force field\cite{AMBER}.
The equations of motion were integrated using the velocity Verlet integrator, with a time step of 0.5 fs.
Long-range electrostatic forces were evaluated using a  particle mesh Ewald algorithm\cite{ewald}. 
A cutoff as large as  9.5 \AA\ was used in  order   to accurately  calculate  the  Van der Waals interactions.  
The atomic partial charges were calculated at DFT level from the electrostatic potential (ESP)\cite{ESP}. 
We then performed finite temperature molecular dynamics on a set of models of increasing dimension containing up to  14400 atoms.
In order to control the temperature we used the Berendsen thermostat using a relaxation time of 0.5 ps.\\

\section{Results and discussion}\label{RES}
\begin{figure}
\includegraphics[scale=0.15]{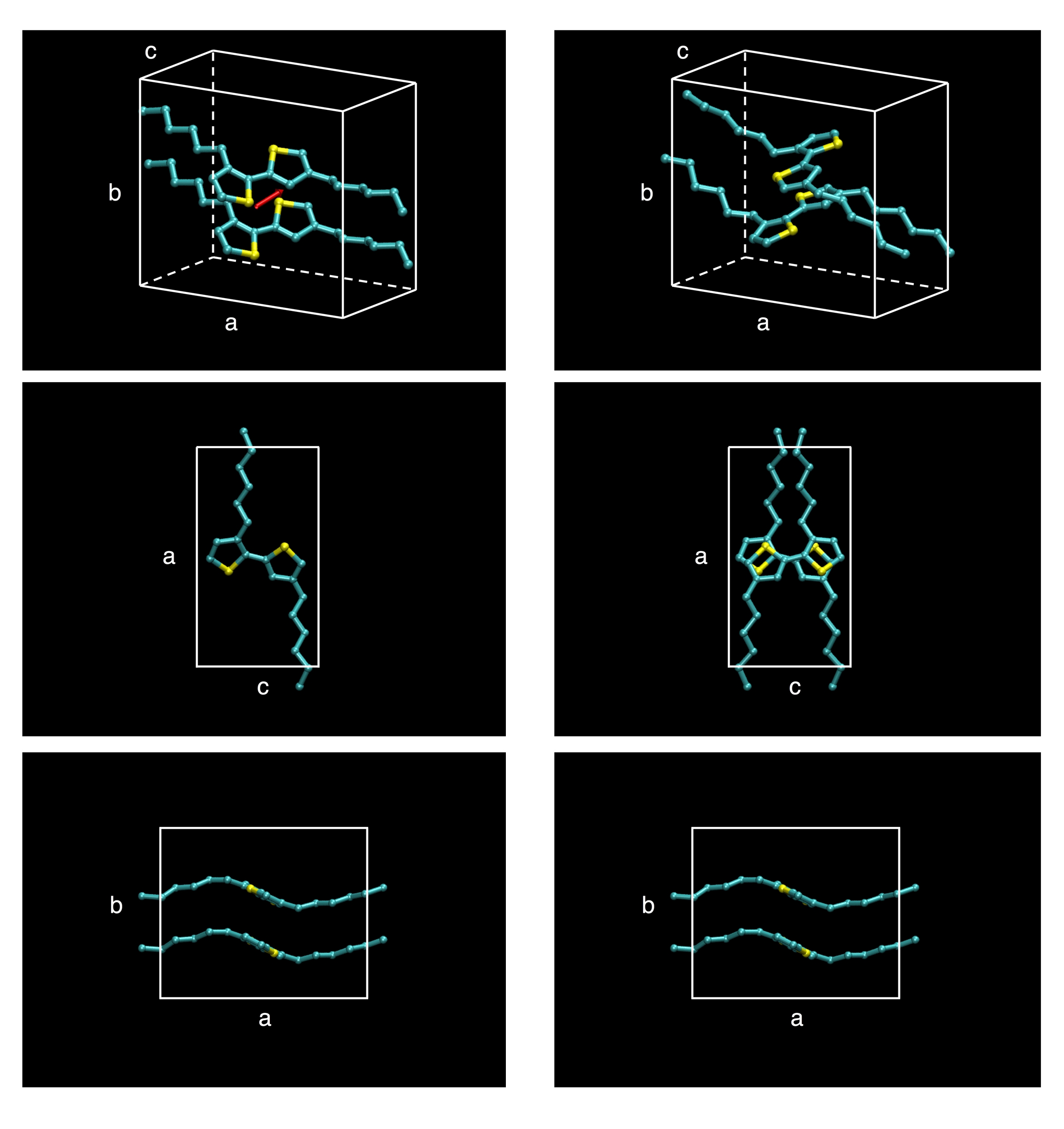}
\caption{  \label{P3HT-STRUCT} Perspective-view (top), top-view (middle) and side-view (bottom)  
of P3HT equilibrium structures obtained by DFT for the aligned (left) and staggered (right)  
configurations. The white box represent the othorombic unit cell with the corresponding lattice parameters. The red arrow 
in the top panel indicates the dipole moment.}
\end{figure}

\begin{figure}
\includegraphics[scale=0.39]{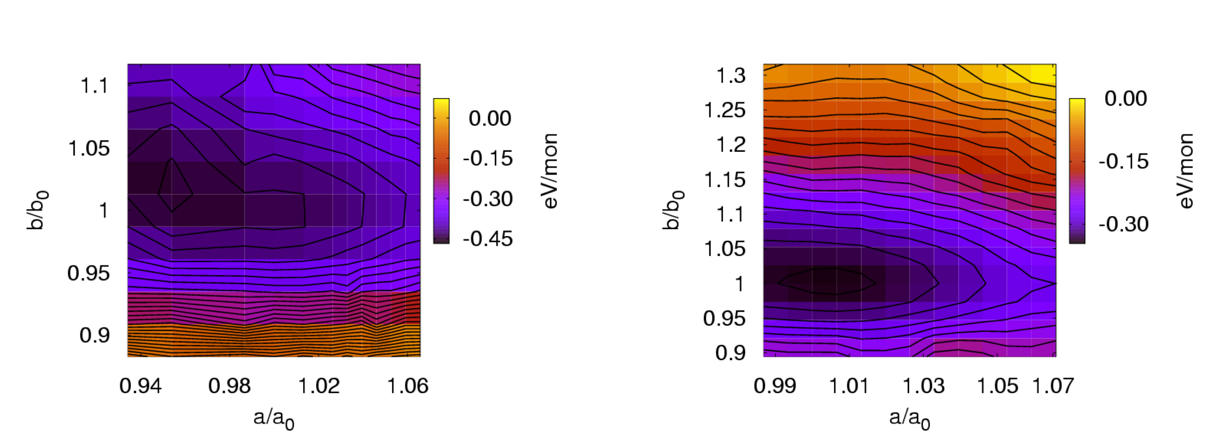}
\caption{  \label{P3HT-MAPS}
Energy landscapes obtained by DFT for the aligned (left) and staggered (right) phase respectively.
The lattice parameters are  referred to the equilibrium values $a_0$ and $b_0$ while the total energy is referred to the energy of two unbound chains.}
\end{figure}

The assembling of two P3HT chains can be driven by two main contributions: the  $\pi-\pi$  interaction which promotes the parallel 
stacking of different
chains  or  the chain interdigitation which promotes the chains  alignment in the same plane.
In order to accurately  compare the two contributions and to identify the major driving force to assembling, we investigated by 
 first principles the  polymer energy dependence on the interchain separation.\\
We considered an orthorhombic unit cell of P3HT with cristallographic vectors  lying respectively in the alkyl side chains ({\itshape a}), stacking ({\itshape b})  and backbone ({\itshape c}) directions (see \ref{P3HT-STRUCT}). Starting from  an ideal planar P3HT crystal structure either in  the aligned  or staggered configurations,
we performed a series of  240 geometry optimizations on  unit cells of different volumes by varying the lattice 
parameters {\itshape a}  and {\itshape b} in the range 14.2-16.2 \AA\ (15.0-16.2 \AA\ for the staggered) and 6.8-10.0 \AA\ both for the aligned and staggered configurations. 
The value of  {\itshape c} was obtained by optimizing an isolated chain and then kept fixed at  7.75 \AA.
 \ref{P3HT-MAPS} shows the corresponding energy landscapes as a function of {\itshape a} and {\itshape b} for the aligned  (left) and staggered (right) configurations. 
In both the staggered and aligned  configurations the energy profile along the {\itshape b} direction (which is the direction where the $\pi-\pi$ interactions occur)
shows a larger variation ($\sim$0.5 eV/monomer ) and a well defined minimum; on the other hand, the energy variation
along {\itshape a} is less pronounced (as small as $\sim$~0.1 ev/monomer), indicating a weak interaction between neighboring chains
due to the interdigitation of the hexyl groups.
This allows us to conclude that the $\pi-\pi$ interactions actually drive the P3HT assembling.

 The aligned  configuration shows two minima. The global one  occurs at  {\itshape a}=14.5 \AA\ and {\itshape b}=7.8 \AA\
 corresponding to a high degree of interdigitation (the interchain distance along the hexyl chains is 14.5 \AA\ ). The thiophene rings  turn out to be inter-twisted along the backbone.
 Such a configuration, though theoretically possible at T=0 K, corresponds to a very high dense P3HT, never  observed  in real systems.
 The second minimum, hereafter named as aligned structure (A), 
falls  closer to the experimental data at {\itshape a}=15.4 \AA\ and {\itshape b}=7.6 \AA. In this case the thiophene rings along each polymer chain lye parallel (see \ref{P3HT-STRUCT}, left panels).

The staggered configurations (\ref{P3HT-MAPS}, right) show only one minimum at {\itshape a}=15.2 \AA\ and {\itshape b}=7.6 \AA\ ,
the corresponding structure having the thiophene rings almost on the same plane (see \ref{P3HT-STRUCT}, right panels).\\
The equilibrium lattice parameters for both the  aligned  and staggered configurations are in good agreement with the 
experimental values: 16 \AA $\leq$ $a$ $\leq$ 16.8 \AA\ , 7.66 \AA $\leq$ $b$ $\leq$ 7.8 \AA\ , 7.7 \AA $\leq$ $c$ $\leq$ 7.8 \AA\ \cite{P3HT-DFT,P3HT-CAST,P3HT-CAST1,P3HT-XRAYS,P3HT-XRAYS2,P3HT-XRAYS3,P3HT-XRAYS4}. 
The largest deviations are found for the  {\itshape a} value as a result of the high interdigitation that is possible only at T=0 K due  to the missing thermal vibrations.
 We will show below that at room temperature the vibrational motion of the hexyl chains hinders the inderdigitation bringing to an average interchain distance of 16.2 \AA\, in agreement with the experimental values. 
A comparison between the aligned  and staggered energy minima shows  that the two configurations are almost isoenergetic (in agreement
with previous calculations), with an energy difference of   0.125 eV/monomer.

The above first-principles analysis is conceptually important in order to identify the interchain interactions. However, because of computational limits,  all the calculations were  performed at T=0 K,  so neglecting the role of thermal disorder in the polymer phase. In order to  include thermodynamics we made use of less computationally demanding MPMD  to perform suitably long ($\sim$ 1 ns) finite temperature simulations on larger systems (up to 10$^4$ atoms).
The accuracy of the  MP requires a fine tuning of the  atomic partial charges  that we achieved by the following  first-principles 
procedure.
We calculated the atomic partial charges at DFT level (see the methodology section) by  embedding
a single P3HT monomer  (two thiophenes with the relative hexyl chains, 50 atoms in total)
in a bulk-like system formed by three P3HT stacked chains (300 atoms) in  the  aligned structure.
We further imposed the symmetry by averaging the charges on the corresponding atoms.
By using the calculated atomic charges, the dipole moments have been calculated for isolated molecules, corresponding to the unit cells at equilibrium reported in  \ref{P3HT-STRUCT}.
The results  show that the dipole moment for the  CPMD-generated A structure is oriented along the molecule backbone (see \ref{P3HT-STRUCT}) with a module of 2.065 Debye, while the dipole moment for the staggered structure is very small (just 0.195 Debye) and oriented along the hexyl chains. 
We stress that the partial charges calculated without the embedding are quite different giving results unsuitable for the bulk.

We validated the above model potential (Amber force field with partial charges calculated as above) by recalculating the energy basins
as a function of the lattice parameters. An overall agreement was found with the DFT results.
Similarly to the DFT case, the MPMD data show two minima for the aligned  configuration and one minimum for the staggered configuration.
The equilibrium lattice parameters are {\itshape a}=15.8 \AA, {\itshape b}=8.0 \AA\ and  {\itshape c}=7.75 \AA\  for the aligned  configuration and {\itshape a}=15.5 \AA ,
 {\itshape b}=8.0 \AA\  and {\itshape c}=7.75 \AA\ for the staggered configuration in good agreement with experiments. Once again, the aligned  and staggered configurations are almost isoenergetic with an energy difference of 0.063 eV/momomer.
\subsection{P3HT assembling}
In this Section we aim at identifying finite temperature realistic models of bulk polymers other than the ideally interdigitated
  structures characterized above. Nevertheless, we find that  the  aligned structure is stable and perfectly interdigitated even after 1 ns-long annealing at T=300K. 

\begin{figure}
\includegraphics[scale=0.089]{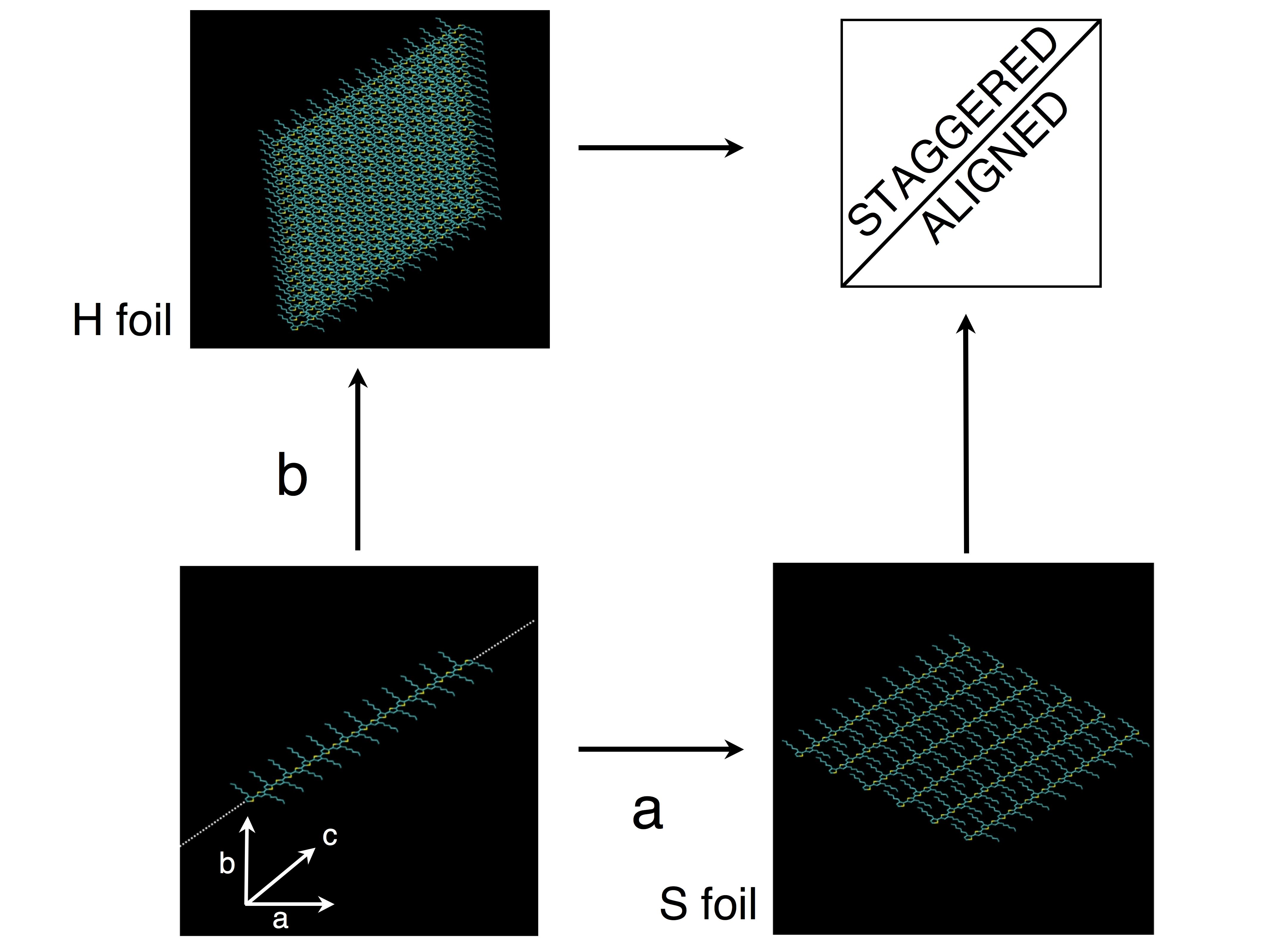}
\caption{Flow chart of the two assembling mechanisms proposed. In the h-mechanism the assembling  of single P3HT chains (lower left) is driven by the $\pi-\pi$ interactions, resulting in the formation of h-foils (upper, left). The successive h-foils assembling gives rise to staggered structures. 
In the s-mechanism the assembling brings to the formation of s-foils (lower right) and their successive stacking gives rise to aligned structures.
 \label{P3HT-ASSEMB1}
}
\end{figure}

Here, in order to find other stable phases, we follow an alternative path and  we focus on the assembling of P3HT chains in vacuo.
The building block of our analysis is a single P3HT chain, extending indefinitely along the backbone direction. 
Firstly we study the stacking of the polymer chains. To this aim, two chains are put at different distances and relative orientations and
  their interaction is studied by  performing 100 ps-long MD runs  at T=300 K.
In the final state, the 
two chains always stacked on top of each other thus confirming that  the $\pi-\pi$ interaction is the driving force for the P3HT assembling.  
Moreover, the staggered configuration was found to be largely favored (and in fact found at the end of any simulation).
By further stacking P3HT chains, two-dimensional structures (hereafter named h-foils) are eventually formed
(see \ref{P3HT-ASSEMB1}).
The h-foils have 
hydrophobic surfaces (h-surfaces)  formed by the hexyl chains.
Moreover, the h-foils give rise to a
 perfectly staggered  stacking.
This is an important result since the direct alignment at T=0 K is the lowest energy configuration for a perfectly interdigitated bulk,
while, for  two-dimensional
structures at room temperature, the staggered configuration is favored.
The above result is confirmed by the fact that,  at  T=300 K,  an  h-foil initially put into an A aligned configuration, always spontaneously  switches to the staggered one.  
Finally, by assembling an increasing number of two-dimensional h-foils  it is possible to generate a bulk polymer
 (see \ref{P3HT-ASSEMB1}).

\begin{figure}
\includegraphics[scale=0.3]{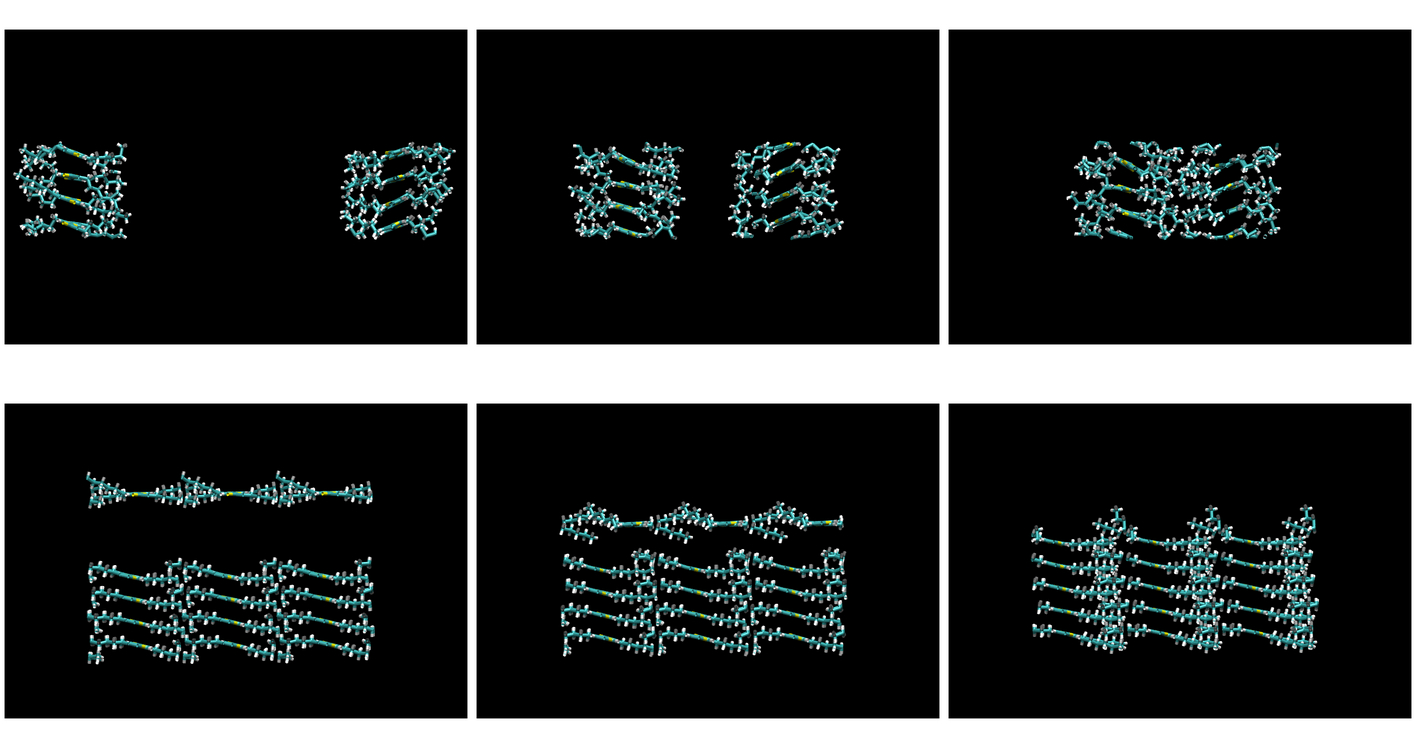}
\caption{Successive MD snapshots corresponding to the assembling mechanisms proposed. In the h-mechanism (top), two h-foils  assemble in a zigzag-like final structure. In the s-mechanism (bottom) one s-foil stacks on top of a P3HT semi bulk in the aligned final structure.
 \label{P3HT-ASSEMB2}
}
\end{figure}

\begin{figure}
\includegraphics[scale=0.4]{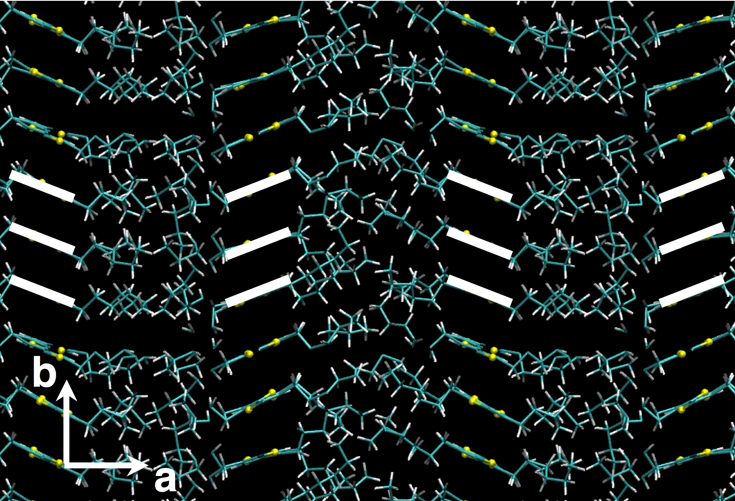}
\caption{Side view of the final zigzag configuration obtained after 1 ns of MD simulation at T=300 K. A zigzag-like structure between the thiophene rings corresponding to different layers is observed.
 \label{P3HT-ASSEMB3}
}\end{figure}

The interaction of the h-foils was studied at finite temperature (see \ref{P3HT-ASSEMB2}, top).  
First of all we put two separated foils at 50 \AA\ distance
and we observed their spontaneous organization into bilayers.
The thiophene rings in the two layers tilt as a result of a long range interaction with the other foil.
 In particular, it was found that the thiophene rings belonging to adjacent h-foils
formed in  a zigzag-like  configuration  (see \ref{P3HT-ASSEMB2}, top). Moreover, this configurations showed a low degree of interchain interdigitation, the equilibrium distance between h-foils was 16.2 \AA\, in agreement with the experimental results.
By replicating the final configuration of \ref{P3HT-ASSEMB2}  a model of P3HT bulk structure was obtained.
We found that the corresponding bulk structure was stable at T=300 K (see \ref{P3HT-ASSEMB3}).
Moreover, the corresponding total energy, obtained  by minimizing the final structure, is  lower than the corresponding total energies for the aligned and shifted configurations by 0.2 eV/monomer.
This crystalline phase  corresponds to the one  previously proposed by Prosa et al. \cite{P3HT-XRAYS}.

This assembling mechanism (hereafter named as "h-mechanism"), ignited by the stacking of two P3HT chains, will occur in absence of any constraint. Moreover, if we suppose the presence of some constraint (like e.g. a substrate) promoting the alignment of two P3HT chains (instead 
of the stacking), the scenario completely changes. In this case we refer to an s-foil the system formed by the alignment of an infinite number of P3HT chains (see \ref{P3HT-ASSEMB1}). In order to investigate the assembling of several s-foils, we simulated the stacking of a single  s-foil on top of an aligned substrate  (see \ref{P3HT-ASSEMB2}, bottom). In this case we found that the s-foil  preserves the 
aligned stacking giving rise to an aligned P3HT crystal. We refer to this assembling mechanism as the s-mechanism (see \ref{P3HT-ASSEMB1}).
\subsection{P3HT surfaces }
\begin{figure}
\includegraphics [scale=1.2]{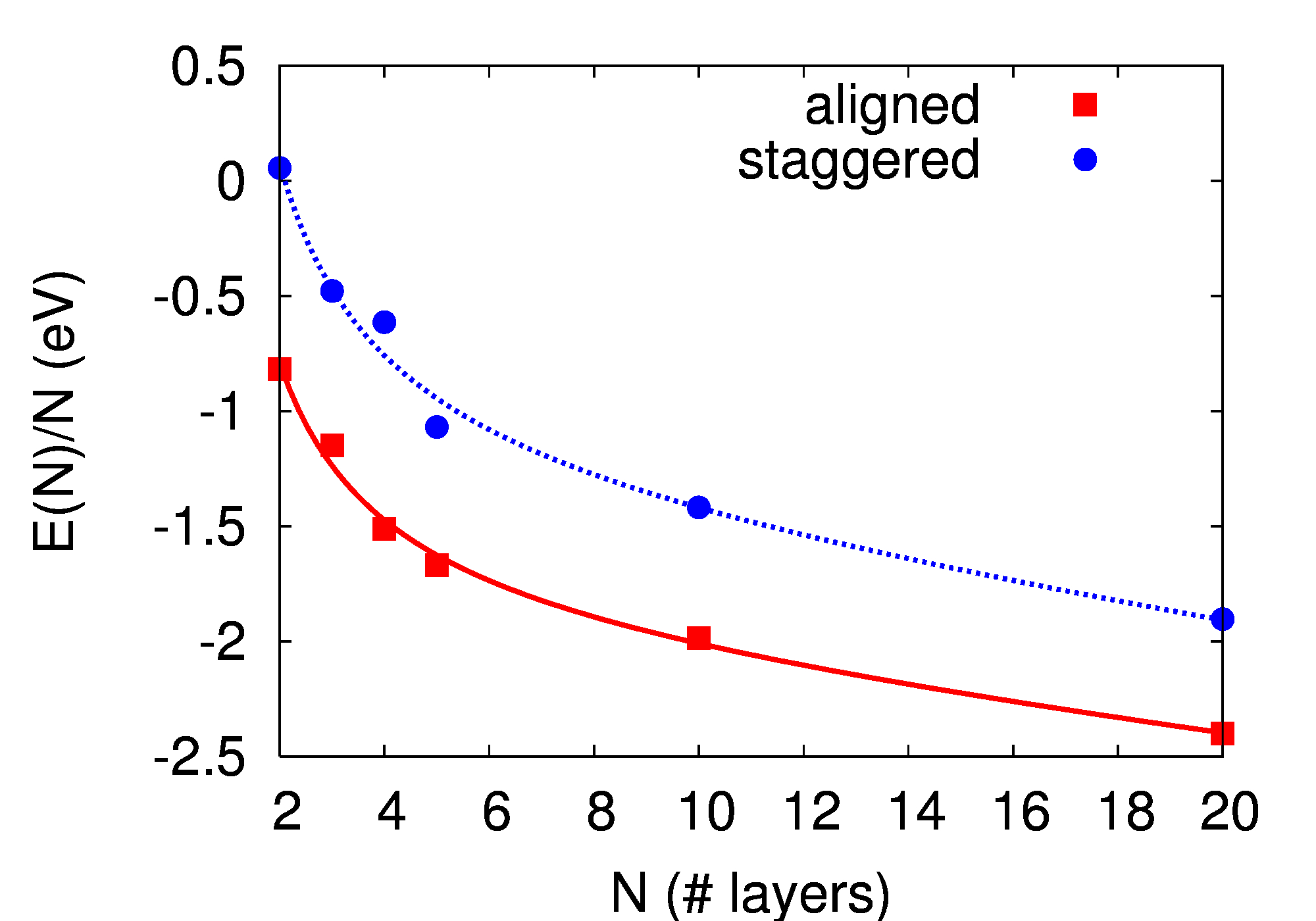}
\caption{ Total energy (normalized to the number of layers $N$) $E(N)/N$ of a P3HT slab formed by a number of layers in the range $2$-$20$.
We considered both the cases  of  initially  staggered  or aligned  configurations.The atomistic data are represented as red squares and blue dots. Lines correspond to the calculations based on the continuum model.
\label{P3HT-SURF}
}
\end{figure}

According to the above analysis, the  h-surfaces are likely to occur in P3HT.
In this Section we calculate their formation energy and their long-range electrostatic interaction.
To this aim, we cut several slabs (formed by an increasing number of layers) out from a polymer bulk, along the h-surfaces.
We considered both staggered or aligned P3HT.
Periodic boundary conditions are applied to each slab by fixing to $15$ nm the distance between the
surface replicas.
After $0.5$ ns-long  MD annealing at low temperature,  
each system was relaxed by  conjugate gradients. 
This procedure is necessary to properly relax the hexyl chains. 
The calculated energies $E(N)/N$ (normalized to the number of layers $N$)  are reported in \ref{P3HT-SURF} as a function of  $N$. 
The energy $E(N)$ of a slab is  the sum of three terms: 
\begin{equation}
E(N) = 2\gamma + E_{B}N-A_{0}N^2
\end{equation}
the energy $2\gamma$ of its two surfaces,  its cohesive bulk energy  $E_{B}N$ ($E_B$ is the bulk energy per layer), and the
electrostatic energy due to the interaction with replicas  that  scales  as $N^{2}$ with the number of atoms of the slab.
$\gamma$, $E_B$ and $A_0$ are used as adjustable parameters to reproduce the atomistic data.
The best fit of atomistic data are reported as lines in \ref{P3HT-SURF}.
  $E_{B}$ was found to be lower in the aligned case ( -6.2 eV ) than in the staggered one (-5.59 eV), consistently with
  the DFT and MPMD results for the bulk.
$\gamma$ was found to be 2.52 eV (3.06 eV)  for the aligned (staggered) h-surface. According to this analysis, 
the assembling of two h-terminated semibulk is more exothermic in the staggered case.
This is consistent with the observed reactivity between staggered  h-foils.
 The $A_{0}$ values are 0.026 eV and 0.033 eV   for the aligned and staggered   h-surfaces, respectively. This means that the staggered h-surfaces have larger interactions than the corresponding aligned one.
\subsection{Conclusions}
In conclusion, we have identified the $\pi-\pi$ as  the main contribution for the P3HT self assembling by means of first principles DFT calculations.
We further studied the P3HT assembling at finite temperature by means of MPMD and we identified  two different assembling mechanism giving 
rise to different final P3HT structures. In absence of constraints, 
the P3HT single chains will stack on h-foils, which at finite temperature are found in the staggered configuration. The successive assembling of
h-foils is predicted to give rise to a bulk structure in a zigzag configuration, as in fact experimentally supported.
Furthermore we studied an alternative assembling mechanism which will occur in  the presence of some constraint (i. e. a substrate) which  favors the alignment of P3HT chain instead of their stacking. In this case we predict the occurrence of an aligned crystal structure.

\begin{acknowledgement}
We acknowledge computational support by CYBERSAR (Cagliari, Italy) and CASPUR (Rome, Italy) and 
Italian Institute of Technology (IIT) under Project SEED "POLYPHEMO". 
\end{acknowledgement}

\bibliography{bibliography}

\end{document}